# Detection and spatial mapping of conductive filaments in metal/oxide/metal cross-point devices

Shimul Kanti Nath[1], Sanjoy Kumar Nandi[1], Shuai Li, Robert Glen Elliman
Department of Electronic Materials Engineering, Research School of Physics and Engineering
The Australian National University, Canberra, ACT 2601, Australia

## Abstract

A simple means of detecting and spatially mapping volatile and nonvolatile conductive filaments in metal/oxide/metal cross-point devices is introduced and its application demonstrated. The technique is based on thermal discolouration of a thin photoresist layer deposited on the top electrode of the cross-point device and relies on the increase in temperature produced by local Joule heating of an underlying conductive filament. Finite element modelling of the temperature distribution and its dependencies shows that the maximum temperature at the top-electrode/photoresist interface is particularly sensitive to the top-electrode thickness. The technique is demonstrated on $NbO_x$ based metal-oxide-metal cross-point devices with a 25 nm thick top (Pt) electrode, where it is used to undertake a statistical analysis of the filament location as a function of device area. This shows that filament formation is heterogeneous. The majority of filaments form preferentially along the top-electrode edge and the fraction of these increases with decreasing device area. Transmission electron microscopy of the top and bottom electrode edges is used to explain this observation and suggests that it is due to a reduction in the effective oxide thickness in this region.

## 1. Introduction

Characteristic resistance changes are observed in two-terminal metal-oxide-metal (MOM) structures subjected to large electric fields or current densities and are of interest as the basis of non-volatile memory and neuromorphic computing devices[1],[2,3]. The resistance changes are commonly mediated by filamentary conduction, either in the form of a semi-permanent filament created by compositional changes in the oxide or as a transient filament created by inhomogeneous current or field distributions (e.g. current bifurcation). Semi-permanent filaments are typically comprised of oxygen vacancies created by the generation, drift and diffusion of oxygen ions during an electroforming step[4], while transient filaments can self-assemble in materials that exhibit S-type negative differential resistance (NDR) due to current bifurcation, a process in which a uniform current divides into domains of high and low current-density [5]. In both cases, the filaments create low resistance paths through the oxide film and can result in significant temperature rises due to local Joule heating when carrying current[6,7].

Knowledge about the structure, composition and spatial distribution of filaments is essential for a full understanding of filamentary resistive-switching and for effective modelling and optimisation of associated devices. As a consequence, there is great interest in methods for detecting and characterising conductive filament. The existence of filamentary conduction is often inferred from area-independent switching characteristics[8] but has also been verified directly by conductive atomic force microscopy (C-AFM)[9] and infrared (thermal) microscopy (IR-M)[7]. Compositional analysis of filaments has been undertaken by transmission electron microscopy (TEM)[10] but this requires pre-knowledge of the filament location and is not suitable for large-scale statistical analysis. Each of these techniques has its advantages and disadvantages but only IR-M is well suited to large-scale statistical analysis of filament distributions and to the detection of transient filaments.

---

[1] Both authors contributed equally.

In this study, we introduce a simple alternative technique for locating volatile or nonvolatile filaments in MOM cross-point devices based on thermal discolouration of a thin photoresist layer deposited on the top electrode of the cross-point device. The technique relies on the increase in temperature at the top-electrode/photoresist interface produced by local Joule heating of an underlying conductive filament and can be used to identify the number and position of filaments. The efficacy of this approach is demonstrated by applying it to filament formation in MOM cross-point devices comprised of a Pt/Nb(Cr)/NbO$_x$/Pt heterostructure.

## 2. Experiments

Pt/Nb(Cr)/NbO$_x$/Pt cross-point test structures, as shown schematically in Fig. 1(a), were fabricated using standard photolithographic processing (supplementary information). Both bottom and top electrodes were defined by a photolithographic lift-off process using negative photoresist (MaN 1420). The bottom electrode consisted of a 5 nm thick Ti wetting-layer and a 25 nm-thick Pt layer deposited subsequently by e-beam evaporation on a thermally oxidized Si (100) wafer with a 200 nm thick oxide layer. A 45 nm NbO$_x$ dielectric layer was then deposited over the entire wafer, including the lithographically defined bottom electrode, using RF sputtering from an Nb$_2$O$_5$ target. Top electrodes, consisting of a 10 nm-thick Nb (or Cr) layer and a 25 nm Pt protective layer, were then deposited by e-beam evaporation. The oxide layer covering the bottom contact pads was then removed by etching to provide direct electrical contact to the pad. The test-devices had electrode dimensions ranging from 2 μm × 2 μm to 20 μm × 20 μm.

As-deposited NbO$_x$ films were analysed by grazing incident-angle X-ray diffraction (GIAXRD) and Rutherford backscattering spectroscopy (RBS) to determine their structure and composition. To simplify the RBS analysis, $NbO_x$ films were simultaneously deposited on vitreous carbon substrates. These analyses showed that the as-deposited films were amorphous, with a stoichiometry of $NbO_x$, where $x=2.60\pm0.5$ (i.e. oxygen rich $Nb_2O_5$). Electrical measurements were performed using an Agilent B1500A semiconductor parameter analyser attached to a Signatone probe station (S-1160). All the measurements were executed at room temperature in atmospheric condition by applying voltage on the top electrode, while the bottom electrode was grounded.

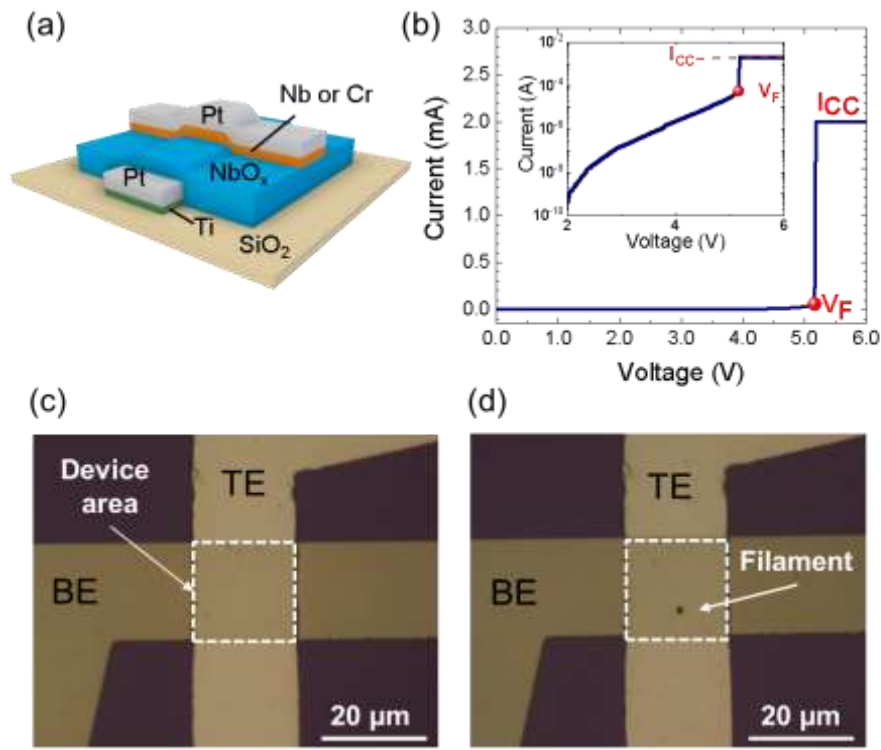

Fig. 1: (a) A schematic of the device structure, (b) a typical current-voltage (I-V) characteristic of the electroforming process for a 20 µm × 20 µm Pt/Cr/NbO$_x$/Pt device. Inset is a semi-log plot of the I-V characteristic. (c-d) optical microscope images of the same device before (c) and after (d) the electroforming process. The dark spot on the device after forming indicates the location of a conducting filament.

### 3. Results and Discussions

As fabricated devices were in highly resistive state (>$10^{10}$ Ω) and required a one-step electroforming process to initiate resistive switching characteristics. This was achieved by applying a positive voltage sweep from 0 V to +6 V to the top electrode while limiting the maximum current through the device (compliance current ($I_{CC}$)) to ~2 mA to avoid permanent damage. Fig. 1(b) shows typical electroforming characteristics for a 20 µm × 20 µm Pt/Cr/NbO$_x$/Pt device. The electroforming voltage, $V_F$ is defined as the point at which the current reaches the compliance limit and corresponds to an electric field of ~2 MV/cm for this devices, which is consistent with previous studies[8]. Following the electroforming step the device resistance is ~10 kΩ, consistent with the formation of a permanent conductive filament.

The high current density associated with filamentary conduction is known to cause a rapid increase in temperature due to local Joule heating. Indeed, in-situ thermal mapping has shown that the surface temperature of the filament can increase by several hundred degrees[7]. Here we exploit this temperature increase to thermally denature a thin photoresist layer and record the location of the filament. To achieve this, the cross-point devices are coated with a thin positive photoresist (AZ 1512HS) layer and a photolithographic step used to reopen access to the device contact pads. The photoresist was then baked for 2 minutes in air using a hot plate kept at a temperature of 358 K (85 °C). Fig. 1(c) shows an image of a 20 µm x 20 µm cross-point before electroforming and Fig. 1(d) shows an image of the same device after electroforming. It is immediately evident that electroforming creates a dark spot in the photoresist, clearly identifying the filament location. Analysis of more than 150 devices showed that forming produced a single filament in most cases; with only a few showing multiple filaments. This is in agreement with previous results indicating the dominance of single filament switching[6, 11, 12].

To determine the relationship between the temperature of the filament and that at the top-electrode/photoresist interface a finite element model of the device structure and filamentary Joule heating was constructed using the COMSOL package. Details of the model are presented in the Supplementary Information. Suffice it to say that the simulation used a 2D-axiosymetric model of the device structure and filamentary conduction process to calculate the steady-state temperature distribution in the device as a function of applied current.

Fig. 2(a) shows the calculated temperature distribution in the active volume of the device for a current of 2.0 mA and clearly shows the temperature increase at the top electrode due to Joule heating of the filament in the underlying $NbO_x$ layer. This is shown more quantitatively in Fig. 2(b) which plots the maximum temperature at the top-electrode/photoresist interface as a function of device current and top (Pt) electrode thickness. This shows that the temperature is a strong function of both parameters and reaches a value of 456 K for a current of 2.0 mA and electrode thickness of 25 nm, more than sufficient to modify the photoresist. However, as shown in Fig. 2(c), the maximum temperature decreases rapidly with increasing top electrode thickness, decreasing from 456 K to 398 K as the electrode thickness increases from 25 nm to 100 nm. Hence, these simulations confirm that the top-electrode/photoresist interface can reach temperatures sufficient to degrade the photoresist but highlight the importance of using thin top electrodes to maximise the sensitivity of the technique. i.e. Filament detection would not be possible in the present case if the electrode thickness was increased significantly.

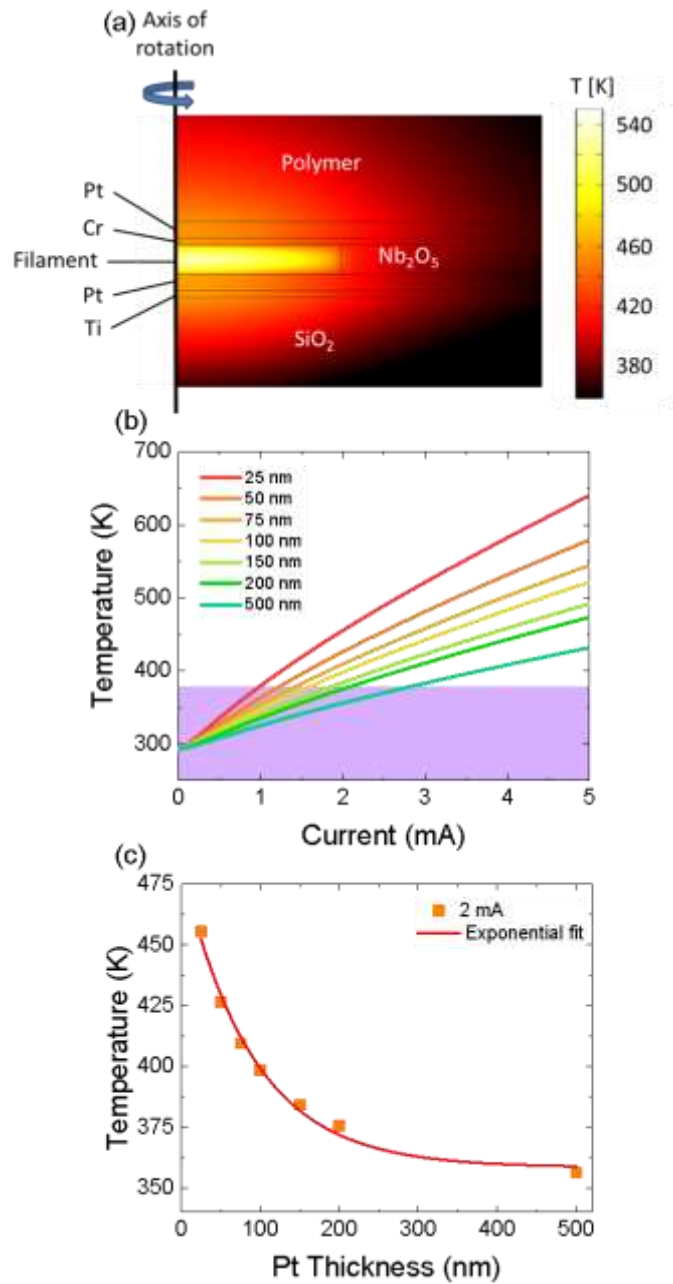

Fig. 2: (a) Temperature distribution in the active volume of the device for a current of 2.0 mA. The depicted structure represents a cross section of an axisymmetric model with the axis of rotation shown. For reference, the thickness of the oxide layer is 45 nm and the width of the conductive filament is 250 nm. (b) Maximum temperature at the top-electrode/photoresist interface as a function of device current and top electrode (Pt)

thickness. (c) Maximum temperature at the top-electrode/photoresist interface as a function of top electrode (Pt) thickness.

The utility of the proposed filament detection method is demonstrated using it to study the influence of device structure on the stochastic nature of electroforming. To this end, the spatial distribution of 60 filaments was recorded for 20 μm x 20 μm (400 μm$^2$) devices using images such as those shown in Fig. 1(d). The results are plotted in Fig. 3(a) and show that a significant fraction of the filaments form at the electrode edges. Moreover, there seems to be a preference for filament formation along the top-electrode edge rather than the bottom electrode edge. To further investigate this point, a similar analysis was performed on devices of different area, and is summarised in Fig. 3(b). This clearly shows that the fraction of filaments formed at the device edges is a strong function of device area, and that there is a stronger preference for filament nucleation at the top-electrode edge for smaller devices; For the 20um x 20 um (400 um$^2$) devices 40% of filaments formed at electrode edges, with ~30% forming at the top electrode (TE) edges and ~10% at the bottom electrode (BE) edges, while for 5 μm x 5 μm devices the edge fraction increases to 75%, with 50% at the top edge and 25% at the bottom edge. These results clearly show that device-related inhomogeneities can play an important role in filament formation. Such effects need to be understood for a full interpretation of area-dependent scaling effects.

For example, the forming voltage for randomly distributed filaments is expected to decrease with increasing device area and to have a dependence of the form[13, 14]:

$$V_f = C_1 - C_2 \ln(\frac{A}{a^3})$$

where $V_f$ is the forming voltage, $C_{1,2}$ are constants associated with statistics of the forming process, $a^3$ is a characteristic material volume, and $A$ is the device area. Fig. 3(c) shows the area dependence for the devices in this study, clearly demonstrating that the forming voltage and dispersion of the forming voltage increase with decreasing area, most likely due to heterogeneous nucleation of filaments at the electrode edges.

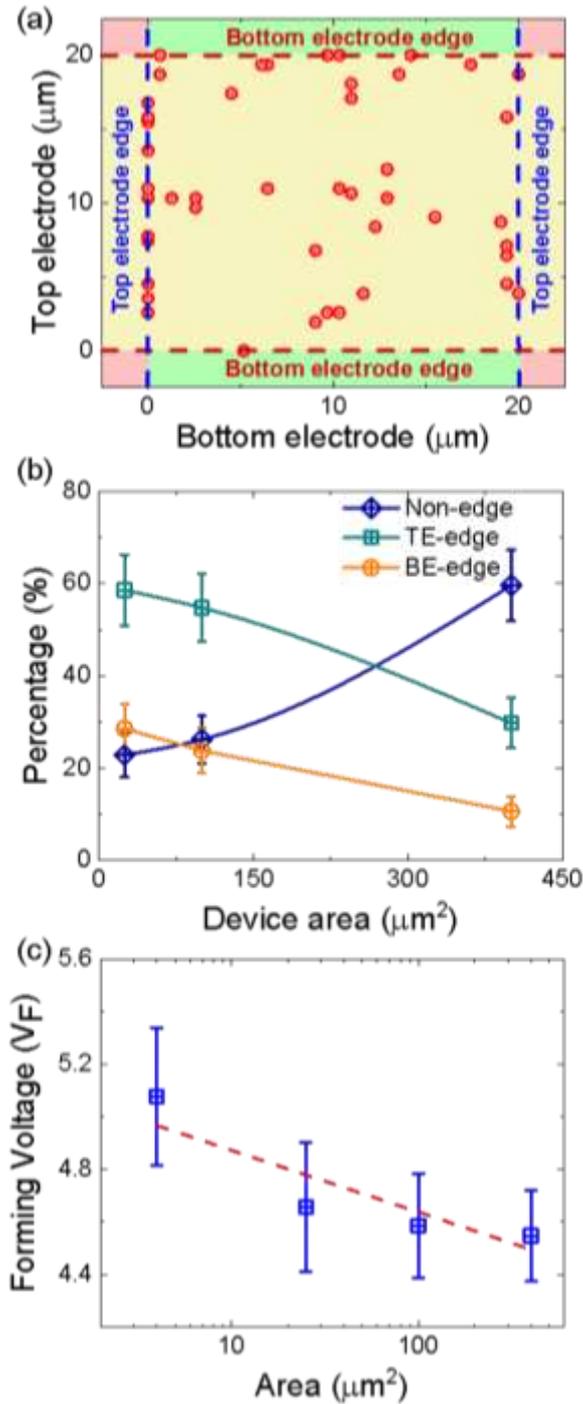

Fig. 3. (a) Location of the filaments of 20 µm×20 µm cross-point, each red dot represents filament in a device. This map was done for 50 different devices. The overlaid dashed lines here show the position of the electrode edges in the cross-point structure. (b) Percentage of devices formed at the edge of the cross-point as a function of device area. (c) Variation of breakdown voltage as a function of device area.

Given that defect formation depends exponentially on local electric field, the probability for forming is expected to increase at the electrode edges, and to be strongly influenced by variations in the oxide thickness, such as those that might be expected for deposition onto the non-planar bottom electrode structure[15]. In this context, preferential forming at the top electrode edge seems counter-intuitive. To better understand how this might arise, cross-sectional TEM images were taken at the top and bottom edges as shown in Fig. 4. These show that the metal electrode layers extend beyond the nominal photolithographic mask edges and into the region where the photoresist is undercut to facilitate the lift-off process, an effect that is clearly evident in Fig. 4(a). The extent of metal penetration increases with increasing metal thickness, so it is greater for the Pt layer than for the Ti, Nb(Cr) wetting layers, and as such produces a region at the edge of the electrodes where Pt is in direct contact with the NbO$_x$ layer, as show in Fig. 4(b). This is significant because the top Nb(Cr) wetting layer is expected to react with the NbO$_x$ to extend the effective oxide thickness. The oxide layer is therefore thinner at the top-electrode edge where it is in direct contact with Pt, possibly explaining why the majority of filaments form in this region.

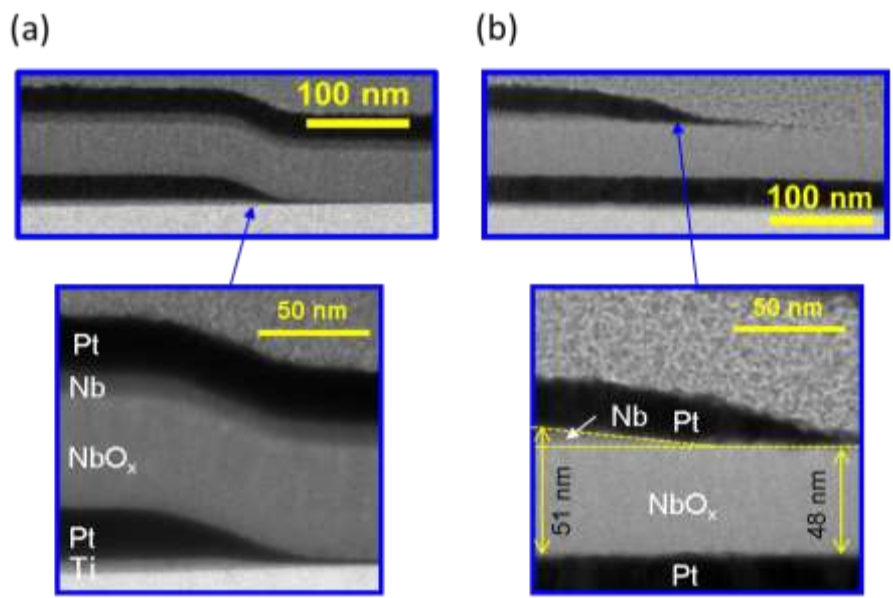

Fig. 4: Cross-sectional TEM of the cross-pint deviceat (a) the bottom edge and (b) top edge.

## 4. Conclusions

In summary, we have demonstrated a simple technique to detect and spatially map volatile and nonvolatile conductive filaments in micron-scale metal/oxide/metal cross-point devices based on thermal discolouration of a thin photoresist layer. Finite element modelling showed that the temperature of the top-electrode/photoresist interface can easily exceed that required to discolour photoresist but that the maximum temperature is a strong function of electrode thickness. The application of this technique to metal-oxide-metal cross-point devices confirmed that electroforming generally created one dominant filament, and showed that filament formation was heterogeneous, with filaments preferentially forming along the edge of the top electrode. TEM analysis of the top and bottom electrode edges suggested that this was due to a reduction of the effective oxide thickness in this region. While further work is required to confirm this hypothesis, the study serves to illustrate the efficacy of the proposed filament mapping technique for understanding such effects. We also note that the sensitivity and spatial resolution of the technique could be improved by using a more temperature-sensitive polymer and further reducing the top electrode thickness.

## 5. Acknowledgements

The authors acknowledge financial support from the Australian Research Council (ARC) Discovery. We would also like to acknowledge the ACT node of the Australian National Fabrication Facility (ANFF) for access to their research equipment and expertise. The authors also acknowledge the facilities, and the scientific and technical assistance, of the Australian Microscopy & Microanalysis Research Facility at the Centre of Advanced Microscopy, The Australian National University.

## **Reference**s

# Detection and spatial mapping of conductive filaments in metal/oxide/metal cross-point devices

Shimul Kanti Nath, Sanjoy Kumar Nandi, Shuai Li, Robert Glen Elliman
Department of Electronic Materials Engineering, Research School of Physics and Engineering, The Australian National University, Canberra, ACT 2601, Australia

## 1S. Device fabrication

Cross-point test structures with dimensions in the range 2 x 2 μm$^2$ to 20 x 20 μm$^2$ were fabricated on a thermally oxidised silicon wafer. Bottom and top electrodes were defined by a standard photolithographic lift-off process, with metal layers deposited by electron-beam evaporation. The bottom electrode consisted of a 10 nm Ti adhesion layer and a 25 nm Pt contact layer, as shown for a 20 x 20 μm$^2$ device in Figure 1S(a). A 45 nm NbO$_x$ layer was then deposited over the entire wafer using RF sputtering from a Nb$_2$O$_5$ target. The cross-point structure was completed by adding the top electrodes, which consisted of a 10 nm Nb (or Cr) adhesion layer and a 25 nm Pt contact layer, as shown in Figure 1S(b). At this point the bottom contact pads remained covered with NbO$_x$ and an etching step was required to access the bottom electrode. This was achieved by defining the area lithographically and chemical etching in buffered hydrofluoric acid to remove the NbO$_x$ layer. Figure 1S(c) shows the cross-point devices after etching the NbO$_x$ layer. The substrate was then ultrasonically cleaned in acetone and isopropanol to complete the device processing. In order to detect filaments, the devices were additionally coated with a thin positive photoresist (AZ 1512HS) layer and a photolithographic step used to reopen access to the device contact pads as shown in Figure 1S(d). Figure 2S(a) shows an SEM image of the cross-point devices and Figures 2S(b-c) show TEM images of the deposited Metal-Oxide-Metal heterostructure.

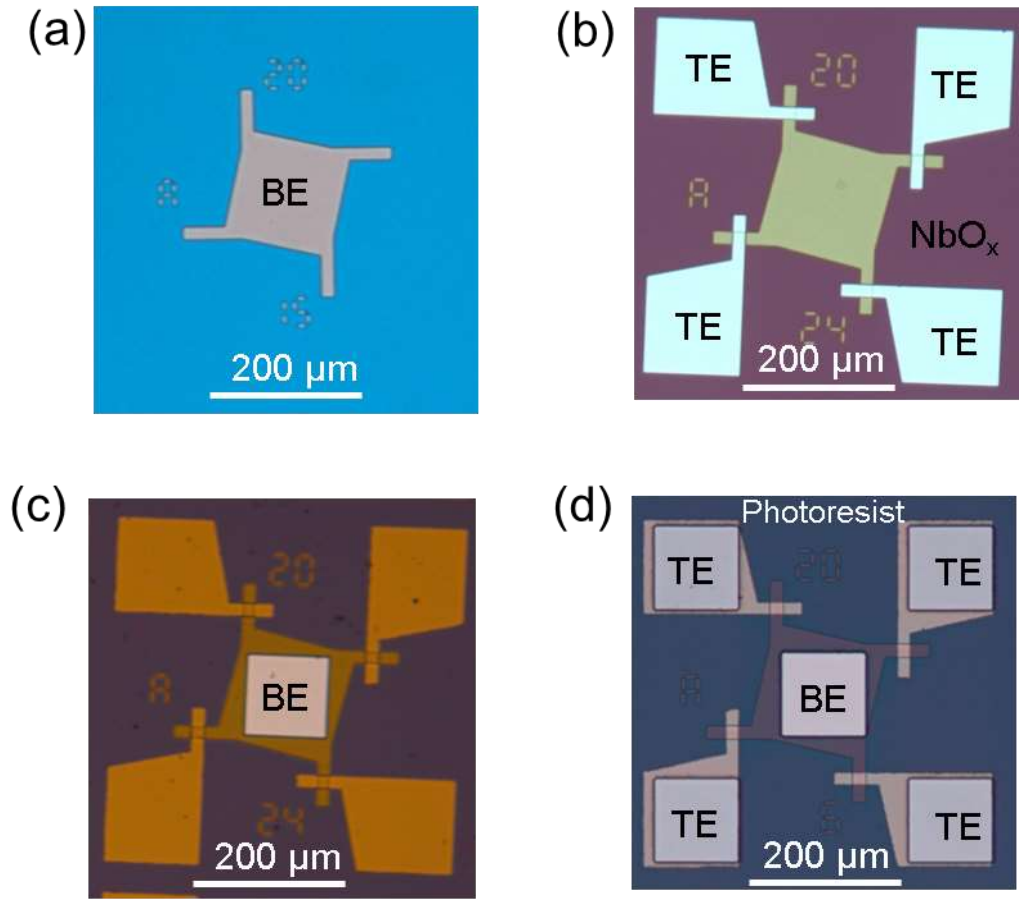

**Figure 1S:** Optical micrograph of different steps of device fabrication: (a) image of a bottom electrode (BE) after the lift-off process for the deposition of a Pt bottom electrode with a Ti adhesion layer. (b) Image of four cross-point devices formed with a common bottom electrode (BE) after oxide and bottom electrode deposition. (c) The same devices after etching the oxide layer to open the BE. (d) Image after the final step, here wafer is coated with photoresist everywhere except the contact pads (BE and TE).

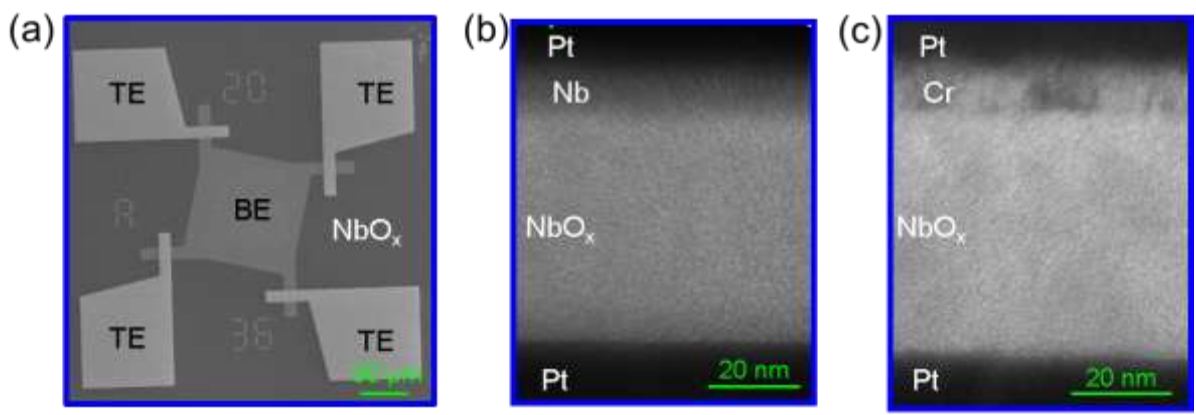

**Figure 2S:** (a) Scanning electron micrograph (SEM) of 20 µm × 20 µm cross-point devices with a common bottom electrode (BE). Cross-sectional transmission micrograph of the device structure: (b) Pt/Nb/Nb$_2$O$_5$/Pt and (c) Pt/Cr/Nb$_2$O$_5$/Pt.

## 2S. Filament distribution

Figure 3S(a-b) show the images of a 10 µm x 10 µm and a 5 µm x 5 µm cross-point devices after electroforming process. The dark regions are showing the location of the filament. The spatial distribution of filaments in 10 µm× 10 µm and 5 µm× 5 µm devices plotted in Figure 3S(c-d).

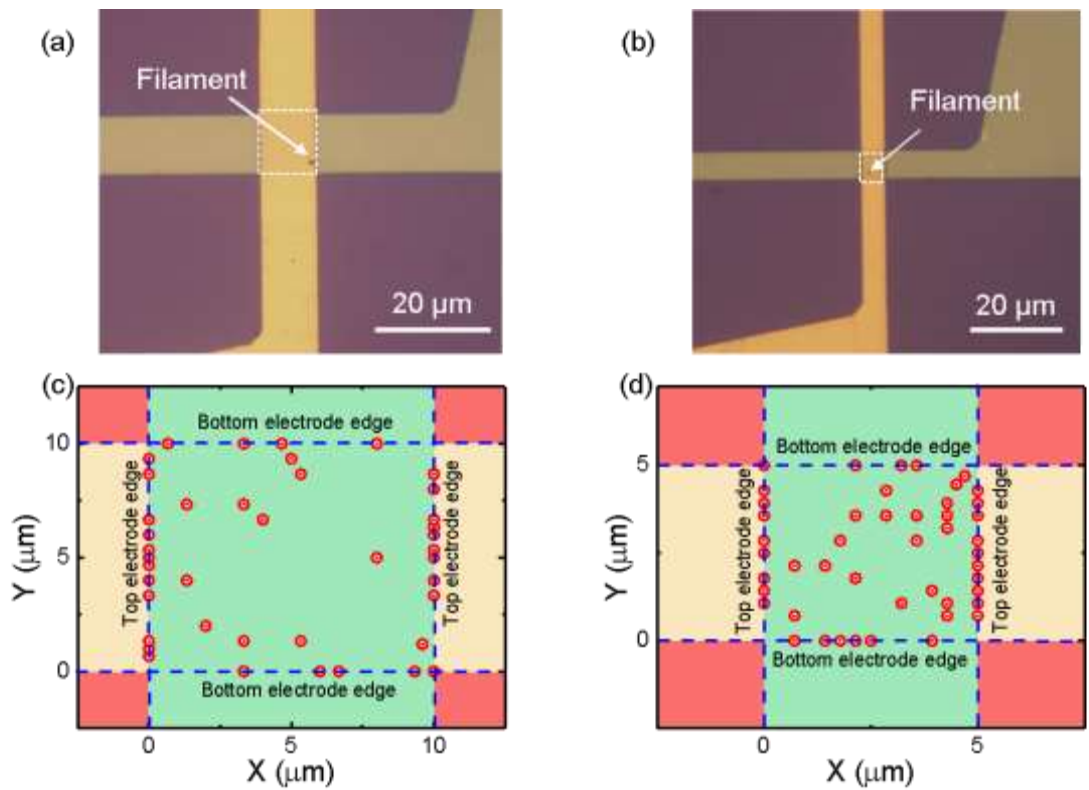

**Figure 3S:** Filament region in (a) 10 µm × 10 µm and (b) 5 µm × 5 µm cross-point devices, and corresponding filament distribution in (c) 10 µm × 10 µm and (d) 5 µm × 5 µm cross-point devices.

## 2S. COMSOL Modelling

A finite-element model was constructed to estimate the temperature rise at the top-electrode/photoresist interface during electroforming. This was undertaken with the COMSOL program using the electric-current and heat-transfer modules to calculate the steady-state temperature distribution as a function of applied current. The device was represented by a 2D axisymmetric model with a device radius of 2 µm and comprised of the following material layers (bottom-to-top): $SiO_2$ (300nm)/Ti (10 nm)/Pt(25 nm)/$Nb_2O_5$ (45 nm)/Cr (10 nm)/Pt (25nm)/PMMA (2000 nm).

The $Nb_2O_5$ layer was assumed to contain a conductive filament of 250 nm radius. Conduction in the conductive filament and the surrounding $Nb_2O_5$ layer was assumed to be due to Poole-Frenkel conduction, as described by equation 1, with an activation energy of $E_a = 0.215\ [eV]$[1]. The absolute conductivities of the two regions was assumed to be different, as determined by the choice of $\sigma_0$. Here we used $\sigma_0 = 2 \times 10^4\ [S/m]$ for the filament and $\sigma_0 = 1\ [S/m]$ for the surrounding oxide layer, so at 293 [K] the conductivity of the filamentary region is ~4.0 [S/m] while that of the surrounding $Nb_2O_5$ is $2 \times 10^{-4}$ [S/m]. The thermal conductivity of the filament and the surrounding $Nb_2O_5$ layer was assumed to be given by $k_{th} = 1.0 + L\sigma T$ [W/(m.K)], where the first term is the lattice contribution and the second the electron contribution. (NB: For the temperatures considered in this study $k_{th} \sim 1.0\ [W/(m.K)]$)[2,3].

The physical properties of the other materials were taken from the COMSOL material library where available. Exceptions were the thermal and electrical conductivities of $SiO_2$ which were set to 1.3 [W/(m.K)] and $1 \times 10^{-10}$ S/m, respectively, and the specific heat and electrical conductivity of PMMA which were set to 1.66 [J/(kg.K)] and $1 \times 10^{-10}$ S/m, respectively.

### 3S.1. Boundary conditions:

The temperature at the top of the PMMA layer and the bottom of the $SiO_2$ layer were set to $T_0$=293 [K], and the lateral edge of the sample was assumed to be thermally insulating. Since Si has a very high thermal conductivity compared to $SiO_2$ its effect was modelled by setting the temperature at the bottom $SiO_2$ layer to $T_0$=293 [K]. However, the temperature distribution at the electrode/PMMA interface was found to depend on PMMA thickness. This dependence diminished with increasing layer thickness and was negligible for layers of 1-2 μm thicknesses, which are typical of the actual layer thicknesses. Simulations were performed for a PMMA thickness of 2 μm.

$$\sigma = \sigma_0 \cdot \exp\left(-\left(E_a - q\sqrt{\frac{q\varepsilon}{\pi\epsilon_0\epsilon_r}}\right)/k_B T\right) - \qquad (1S)$$

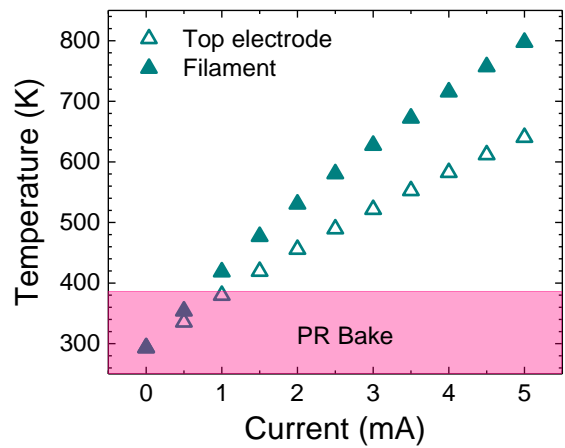

**Figure 4S:** Maximum temperature in $Nb_2O_5$ filament compared to maximum temperature at top-electrode/PMMA interface

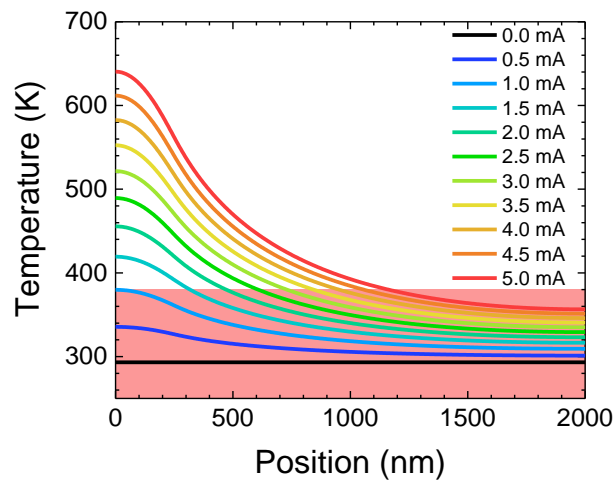

Figure 5S: Temperature distribution as a function of distance from the centre of the filament as a function of current.

## 4S. Edge effect

To further clarify the edge effect, additional cross-sectional TEM images of the bottom electrode of a $Pt/HfO_2/Pt/Ti/SiO_2/Si$ structure are demonstrated in Figure 6S. This clearly shows that the bottom Pt layer extends beyond the Ti adhesion layer in the undercut region.

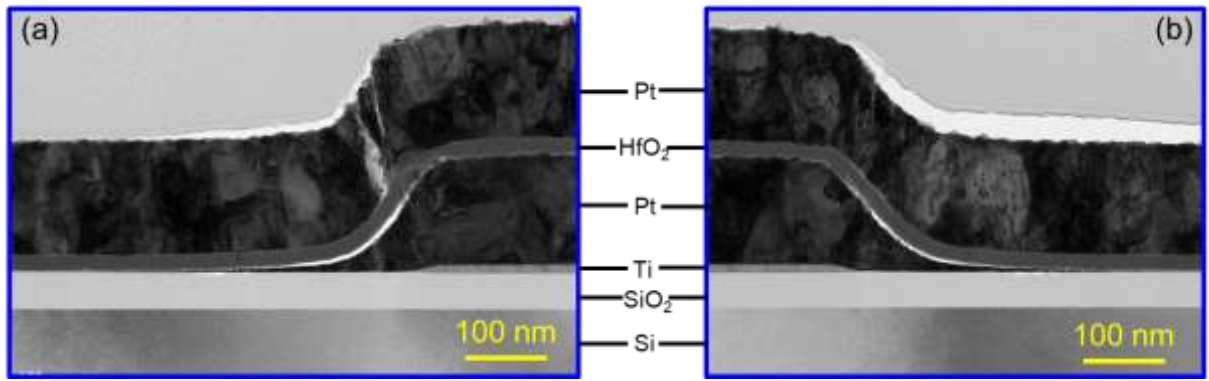

**Figure 6S:** The undercut effect in Pt/HfO$_2$/Pt/Ti/SiO$_2$/Si str5ucture show left and right edges of bottom electrode.